# Navigating Ethics and Power Dynamics through Participant-Designer Journey Mapping

Leonor Tejo
*DEI — CISUC*
*University of Coimbra*
Coimbra, Portugal
0009-0001-8749-2866

Paula Alexandra Silva
*DEI — CISUC*
*University of Coimbra*
Coimbra, Portugal
0000-0003-1573-7446

*Abstract*—As Digital Transformation and innovation driven by Information and Communication Technology (ICT) continue to mark the evolution of society, ethics emerges as a central concern not only in terms of the outcomes and implications of technological systems but also throughout the activities carried out through the development of those systems. Power dynamics have been identified as a recurrent ethical challenge in the design and development process. As designers, participants, and project stakeholders engage in the process, potential conflicts, power imbalances, and ethical challenges emerge. This requires that awareness is raised on these imbalances and that teams proactively act on them. To address this issue, we propose the Participant-Designer Journey Map (PDJM), a tool to assist designers in conducting an ethical design process, aware of power imbalances. The proposal for the PDJM was evaluated, based on a set of criteria derived from the literature, by three design professionals, against nine other alternative tools. The PDJM was identified as the tool with the most potential to facilitate a structured approach to navigating ethical dilemmas, particularly those related to power dynamics.

## I. INTRODUCTION

Our daily lives are increasingly interconnected with technology, and the way we interact with it is evolving at a notable speed [1]. This shift has major implications for the field of Human-Computer Interaction (HCI) and the methods used by professionals to create interactive artifacts [2], [3], such as those leading to digital transformation.

However, innovation is not a panacea. Designers and other tech-related professionals are increasingly aware of the ethical dilemmas surrounding the creation of technological artifacts, as stressed in the work of Malik and Malik [4] that describes the 'awakenings' among practitioners towards social change.

In recent years, there has been an effort to understand how designers understand ethics and apply it during the design process [5]–[11]. In this context, one challenge – power dynamics – emerges as central in the design process, underlining the role that power plays in the design process [5].

We have studied ethical concepts and existing approaches in HCI (e.g., [1], [12], [13]) and contemporary tools (e.g., [14], [15]) to better understand existing solutions and their objectives [16]. In addition, we have analyzed how power dynamics are experienced and what strategies are applied to address them.

Power dynamics have significant implications for the ethical design process, as they are influenced by decision-makers and those in positions of authority, and can impact how participants are represented in the project [17]. While power dynamics are inevitable when using participatory approaches [5], [18], power dynamics are not desirable and can have a negative impact on the design process [5], [17], [18].

Power imbalances, where designers have the most influence and control, can compromise the ethics of the project [18]. This can also undermine the authenticity of the participation of those involved and reduce the potential for positive impact on the project [17], [18]. In corporate, and sometimes even in research settings, further systemic imbalances may occur among the technical staff, the project management team, and corporate management that affect power [5], [8], [9]. While we acknowledge this issue, our work has a particular focus on the power imbalances that may emerge between designers and participants.

Based on our developed understanding of the power imbalances between designers and participants, namely on participatory design approaches, this work contributes the Participant-Designer Journey Map (PDJM), a tool that documents the collaboration between designers and project stakeholders with a focus on end-user participants, identifying points of interaction and highlighting power imbalances. While our method strives to involve all relevant stakeholders, from the technical staff to management and participant stakeholders, we do not directly look at how to deal with potential instances where middle and/or top management may overrule decisions made by the designer guiding the mapping process or by those directly interacting with end-user participants.

The PDJM was evaluated against nine other alternative tools, based on a set of literature-based criteria, by three design professionals, and was identified as the tool with the greatest potential to address conflicts and power imbalances, while also improving ethical and responsible innovation, which is at the core of the design process.

## II. EXISTING THEORIES AND PREVIOUS WORK

### A. Ethical Challenges and Power Dynamics in HCI

The concept of power is described as the "structuring dominance and order in organizations", showing that power is "not simply a relationship between partners, individual or collective; it is a way in which certain actions modify others" [17]. When adopting a participatory design approach, the dynamics of power in the design process can become a central issue, despite the fact that the approach was created to address power imbalances [18].

Problems arise primarily during the decision-making process, where designers are typically responsible for the final decision [17]. This decision-making process refers to both who take part in the design process and how they will be represented in the final decision [18]. In contrast, the power of the participants is perceived as being "indirect" and essentially being carried out through feedback exercises or the generation of knowledge for the project [17]. This issue is emphasized due to the fact that participants perceive the designers as having more power and do not know how to contribute effectively and express their needs, which limits their participation [17], [18] and their potential influence on the project [17]. Once a participatory approach is adopted, the complexities of such an approach have to be recognized and designers should be aware of the aspects, e.g. language or cultural factors, that influence power dynamics [19].

Analyzing recent studies on how HCI designers understand ethics and engage with ethics in the professional environment [5], [7]–[11], the theme of "power" emerges predominantly as the source of challenges in the design process. These power dynamics are described between various project stakeholders, including between designer and participant [5], leadership positions and designer [5]–[11], between designers [5]–[9] and between participants [5]. The challenges related to power dynamics between leadership positions and designers and between designers are mostly described as resulting in a lack of decision-making power [5], [8]–[10]. These challenges are due to hierarchies within the company [5], [10], hierarchies within the teams themselves, particularly due to the level of experience [9], or due to the goals of the organization, or client [5], [10]. Another challenge is the presence of hierarchies between designers and participants [5], showing that this heterogeneous relationship affects the partnership that should be built between the two project parties. These plus the desire to change hierarchies within organizations [10] and of developing a greater awareness of power dynamics by designers [5] further motivate this study.

### B. Strategies and Methods to Support Ethical Design in HCI

A strategy to deal with ethics in design is through an open and inclusive dialogue with all the people involved in the project [5], [17], [20]–[24]. From understanding the context [20] to defining what is considered "harm" [5], [23], it is argued that there must be communication and understanding between all parties involved in the project on all aspects of the development. Additionally, both the designer and the participant should be aware of the values and interdependence of their interaction and relationship [5], [22]. While studies underline that those in power must be aware of the different biases or preconceived ideas they may have [19], [21], [24], others mention the need to hand over power to the participant [19]. Awareness of the power systems in place [21] and the importance of identifying "points of resistance" is also emphasized [20]. Finally, the need to fully understand the context and for whom the design object is being made is recognized, making it necessary to create moments of reflection before decision-making [19], [21].

Several tools have emerged to support ethical design, demonstrating the growing interest in responsible design among HCI and design professionals [16]. These methods address a range of goals, including anticipating the impact of technology [25], and creating active reflection on ethics during the design [15]. Chivukula et al. [16] examined 63 ethics-centered methods to analyze their application, intended use, audience, context, and structural elements. The authors discuss methodologies that cover a wide range of subjects, from social constructions to privacy, and are often used in conjunction with frameworks such as Virtue Ethics or Value Sensitive Design.

The reviewed tools provide ethical guidance in various formats, such as steps, guidelines, frameworks, perspectives, reflective questions, examples, heuristics, and case studies, presented in worksheets, templates, cards, documents or guidebooks, videos, and games [16].

However, despite the variety of methods, only one tool has been found that addresses power dynamics: the Data Ethics Canvas [14]. This tool is specifically designed for "anyone who collects, shares, or uses data" [14], with a focus on data ethics. This tool seeks to facilitate transparency between designers, stakeholders, and participants by bringing together different project parties to make and reflect on decisions. While this tool represents a positive step towards addressing power dynamics, namely by addressing aspects of transparency in data usage and sharing, it does not explicitly address the relationship between designers and the various stakeholders involved in a project throughout the design process. Several different issues, other than those related with data, may emerge across the development of a project, which can raise ethical concerns, so we need methods that extend beyond the Data Ethics Canvas, namely by contributing to the identification of stakeholders' values and expectations, the mapping of potential points of power imbalance and the definition of strategies to deal and reflect on those imbalances.

## III. METHODS

This study aims to create a tool to support design and development teams fostering an ethical, responsible, and power-balanced process throughout a project. In addressing this goal, this study poses the following research question: How can design teams be supported in following an ethical design process that maintains a power balanced relationship among project stakeholders, in particular end-user participants?

To address this goal, and building upon the insights gathered from previous work, the two authors, who have a background in design and multimedia and in HCI, set off to generate ideas for tools which could help address power dynamics. These ideas of alternative tools were then analyzed against a set of criteria described in the scientific literature and assessed for their potential by three design experts from the HCI field. The tool identified as the one holding the most potential, the PDJM, was then fully developed and documented, making up for the main contribution of this paper. The following sections describe the process followed to create the PDJM.

### A. Ideation

For the ideation process, we first created a problem statement that captured what the tool was intended for: "Designers who want to follow a power-balanced design process need an integrated, power-aware tool that supports and guides them in conducting an ethical design process". This statement covered: who the tool is being designed for — "Designers who want to follow a power-balanced design process" — what the characteristics of the tool are — "seamless, power-aware" — and its purpose — "to support and guide them in conducting an ethical design process.".

With the problem statement in mind, the two authors used the 6-8-5 method [26] to generate ideas for alternative tools that could address the problem statement. This method encourages the exploration of a large number of alternative ideas by generating six to eight ideas in successive rounds of five minutes. Ideas are captured on a sheet of paper with a grid-like structure, created by having a central vertical line dividing the page into two halves, and three horizontal lines intersecting the central line. In filling out the sections, the goal is to generate as many ideas as possible without the need to consider the details of any of them [26].

This procedure resulted in the generation of thirty-two ideas for potential tools that were further developed and elaborated upon to flesh out how the idea could effectively support teams in conducting an ethical design process. The thirty-two ideas were finally organized into a table that consistently described each idea in greater detail regarding the format and general procedures of the tool.

### B. Selection

Following idea and concept generation, we focused on selecting the most appropriate ones. Based on the literature review on power dynamics in the design process, we developed a set of 10 criteria and captured them on a list (Table I), that was then used to support the evaluation and selection of alternatives.

Table I shows the list of criteria and their corresponding definitions, informed by the literature. The criteria included aspects such as what a tool should provide, what characteristics it should incorporate, and how it could be integrated in the design process. Tejo [5] states that ethics should be reviewed at more than one stage of the project, so the tool should also encourage this review throughout the design process. Finally, the tool should be easily integrated into existing design processes and should not require too much effort on the part of the team. While the list privileges criteria that directly address issues of power dynamics, other aspects, such as how usable and learnable the tool would be, are covered in other all encompassing criteria, in particular the tool being seamless to apply and adaptable.

TABLE I
DEVELOPED CRITERIA FOR DESIGN PROPOSAL EVALUATION.

| Criteria | Definition |
| --- | --- |
| Seamless | The tool should be easy to integrate into the design process which is already in place [5]. |
| Manageable effort | The tool should require limited dedicated time from the team [5]. |
| Adaptable | The tool should adjust well to different projects [5]. |
| Pedagogical | The tool should help to identify and make designers become increasingly aware of different power dynamics, as well as possible biases or preconceptions they may have [5], [19]–[22], [24]. |
| Proactive | The tool should help the team foresee and prevent potential power dynamics issues that may arise in the project [5]. |
| Obstacles identification | The tool should facilitate the recognition of power systems and those who holds power, enabling the identification of points of resistance [20], [21]. |
| Inclusive dialogue | The tool should actively involve all project stakeholders (the participants and the team itself), and encourage the discussion of the perspectives of different stakeholders [5], [17], [20]–[24]. |
| Mutual dependency | The tool should encourage a true partnership between researchers and project stakeholders, recognizing the role and importance of each [5], [19], [22]. |
| Reflective action | The tool should encourage a continuous process of reflection that translates into action within the project [19], [21]. |
| Sustainable impact | The tool should incentivize the continued follow-up of an ethical, power-oriented process [5]. |

The list of criteria was then used to make a first selection of the most appropriate tools. Only the tools that fulfilled seven or more criteria were kept, resulting in ten tool alternatives.

### C. Evaluation by experts

The selected proposals were systematized in a card format (Figure 1 show two examples of those cards) that could be used to describe the 10 alternative tools in a coherent fashion. These 10 alternatives were then reviewed by three HCI practitioners from three different professional contexts — academia, applied research centers and corporate.

The evaluation was conducted using an online questionnaire, where the experts scored each proposal using a 5-point Likert scale against each criterion. At the end of the questionnaire, two open-ended questions were asked to find out the subjective opinion of the participants, without taking the criteria into account. The first question asked for a subjective preference of a maximum of three proposals, while the second asked if any of the tools they mentioned were developed, which one(s) they would apply in their projects. The participants were also provided with a document explaining each criterion and advised to keep it open while reviewing the proposals.

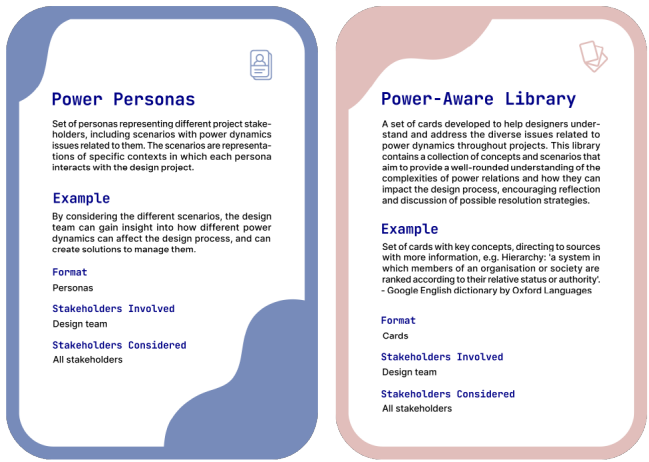

Fig. 1. Example of the selected proposals' cards.

For the selection process, each criterion was given a score on the Likert scale, with "strongly agree" corresponding to five points and "strongly disagree" to one point. According to this system, the score for each design proposal was calculated. The three highest-scoring proposals were then ranked. In addition to the distribution mentioned above, one of the proposals stood out in the responses to the open-ended questions. Two participants stated that if that proposal were created, they would apply it to their projects, while one participant expressed a subjective preference for that proposal. This information was taken into account and used to establish a tie-breaker.

The tool selected for development was the Participant-Designer Journey Map (PDJM), which fulfilled all the criteria defined and was preferred by the designers who evaluated the proposals. Table II reviews how the tool meets the criteria. The PDJM consists of a visual representation of the designer's and participants' joint journey through the design process. The tool maps the different points of interaction between designers and parties and allows for the identification of possible moments of tension, as well as opportunities for the empowerment of the participant. Through this map, it is possible to anticipate problems and take steps to resolve them before they happen.

## IV. Results

The Participant-Designer Journey Map (PDJM) is a tool designed to understand, map, reflect and document the collaboration between designers and participants across the design process. It focuses on identifying points of interaction and tension to anticipate potential issues and plan strategies to resolve conflicts as they arise, and implement corrective measures to promote effective collaboration. It is crucial that the design team understands that the main purpose of the tool is to address power imbalances and encourage open dialogue between all parties involved in the project. All mapping should be conducted in a collaborative manner, with the objective of ensuring that all project stakeholders are fully cognizant of the project dynamics and their respective roles within the project, including designers, participants, technical staff, and other stakeholders. The PDJM should evolve along the project and with each new project. In this way, the tool is not intended for single use, but rather to promote continuous dialogue and power balance throughout the design process.

TABLE II
HOW THE PARTICIPANT-DESIGNER JOURNEY MAP (PDJM) MEETS CRITERIA.

| | |
|---|---|
| Seamless | The PDJM can be seamlessly integrated into the existing design process, providing an additional layer that increases the team's power dynamics awareness. |
| Manageable effort | The PDJM does not require significant additional time or resources and does not disrupt the normal workflow. |
| Adaptable | The PDJM is adaptable to different projects, allowing adaptation to the unique contextual dynamics of each project. |
| Pedagogical | By visually representing the Participant-Designer journey, the PDJM helps designers identify and understand power dynamics, biases and preconceived ideas, fostering learning throughout the project. |
| Proactive | The PDJM supports the team in anticipating potential issues of power dynamics enabling the adoption of proactive measures. |
| Obstacles identification | Through visual representation, the PDJM helps recognising power systems, identifying who holds power and points of resistance. |
| Inclusive dialogue | The PDJM allows for an open debate with all project participants, ensuring that diverse perspectives are considered through collaborative decision-making. |
| Mutual dependency | By illustrating the journey from the perspective of both the designer and the participant, the PDJM recognizes and reinforces the interdependence between researchers and the various stakeholders. |
| Reflective action | The PDJM encourages reflection at each stage of the journey, leading to informed actions, aligned with the team's ethics. |
| Sustainable impact | By using the tool, it is possible to learn and consequently integrate the knowledge acquired into new projects. |

The tool is composed of a journey map template[1] (Figure 3) and a guide[2] (Figure 2) that helps to understand each field and provides suggestions on how to use it. The journey map contains three main sections: i) Understanding, Mapping, and Reflecting. Each section contains fields, that we refer to as swimlanes, that capture essential information to understand the relationship between designer and participant, decision-making processes, and power dynamics; these are required swimlanes. There are also optional swimlanes that afford flexibility for further mappings according to the context and needs of the project at hand.

---
[1]PDJM use guide: https://e4ay.short.gy/GuidePDJMpdf
[2]PDJM template: https://e4ay.short.gy/PDJMpdf

To guide the team along the design process, the journey map includes the main phases of the double diamond design process [3]. In addition, the PDJM includes an alignment phase that aims to establish a solid basis for collaboration by identifying shared values, methods of power distribution, and expectations. The alignment phase also serves to discuss approaches that could be applied to the design process and to identify negative and external influences that might affect the project, e.g., budgets, regulations, and industry trends.

To illustrate the use of the PDJM, we will explain how to fill-out the different sections and swimlanes of the journey map using a fictional example of a project aimed at developing an application to support gymnasts training. This fictional project seeks to support the training of pirouettes by introducing a pair of smart insoles to analyze movement and improve performance. Power dynamics are a risk in this project, since the gymnastics training environment inherently presents hierarchies and the main designer leading the project is also a coach at the gymnastics academy where the study is being carried out. In addition, introducing technology in an environment where technology is scarce to non-existent can raise further challenges. Table III describes each section and swimlane and how to fill out each.

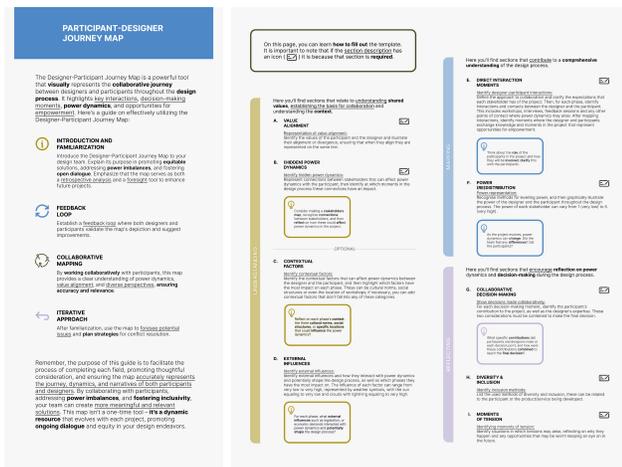

Fig. 2. PDJM use guide, available at https://e4ay.short.gy/GuidePDJMpdf.

### A. Understanding

The swimlanes in the 'Understanding' encourage an understanding and discussion of shared values, as well as an understanding of the context considering various project stakeholders, while also encouraging a deeper understanding of how factors outside the project can affect power dynamics.

[3]https://www.designcouncil.org.uk/our-resources/the-double-diamond/.

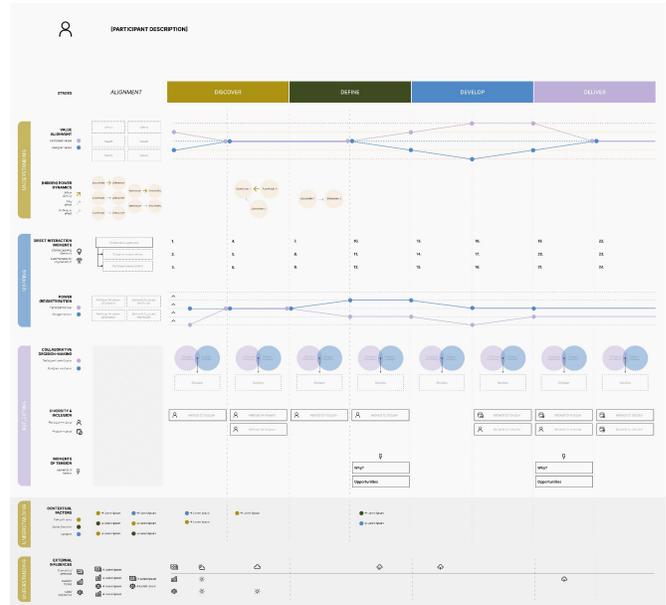

Fig. 3. PDJM template, available at https://e4ay.short.gy/PDJMpdf.

*1)* **Value alignment**: This swimlane has a moment of alignment which consists of identifying the values of the designer and the participant; throughout the project, these values are illustrated graphically as they converge or diverge.

Considering our example, during the alignment of values phase, the goals of "efficiency", "autonomy" and "safety" were identified for the project (Figure 4). The mapping shows that, at the start of the project, the values of the designers and participants were not aligned, requiring an initial negotiation process. Then, for most of the project, the values were aligned, except from the 'define' phase, when there was a misalignment, which was later realigned in the 'deliver' phase. Figures 4-*External Influences*, 5-*Power (re)distribution* and 6-*Moments of tension* below show that this misalignment is attributed to a moment of tension, potentially caused by external influences.

The Value alignment swimlane facilitates the identification and representation of shared values between participants and designers. It also helps create a basis for collaboration by ensuring a common understanding of core values.

*2)* **(Hidden) power dynamics**: The mapping of (Hidden) power dynamics focuses on the relationships between stakeholders that may affect the power dynamics with the participant, by first identifying them and then positioning them at the various moments of the design process. Figure 4 shows the relevant relationships for our exemplar project, with a significant impact on power dynamics. For example, the relationship between "coaches" and "athlete" carries considerable significance, as the hierarchies inherent in the relationship between these two stakeholders can affect the project. In addition, if one considers that the main designer leading this project is also a coach where the study is being carried out, their influence on the power of the participants (athletes) is very strong. In the 'define' phase, these relationships had an

TABLE III
EXPLANATION OF EACH SECTION AND SWIMLANE OF THE PARTICIPANT-DESIGNER JOURNEY MAP.

| | Name | Alignment Phase | Design Process Phases |
|---|---|---|---|
| **Understanding** | Value alignment | Consists of identifying the values of the designer and the participant | These values are illustrated visually as they converge or diverge, ensuring that when they align they are represented by overlapping lines |
| | (Hidden) power dynamics | Represents relationships between stakeholders that can affect power dynamics | Identify at which moments in the design process these connections have an impact |
| | Contextual factors* | Involves identifying contextual factors (e.g., cultural norms, social structures, or workshop location) that may impact power dynamics between the designer and the participant | Highlight which factors have the most impact on each phase |
| | External influences* | Captures external influences, such as economic constraints or regulations | Identify in which phases external influences have the most impact; the influence of each factor can range from very low to very high, represented by weather symbols, with very low represented by the sun and very high by lightning clouds |
| **Mapping** | Direct interaction moments | Defines the approach to collaboration and clarifies the expectations that each stakeholder has of the project | Identify interactions and contacts between the designer and the participant; after mapping interactions, identify moments where the designer and participants exchange knowledge and moments in the project that represent opportunities for empowerment |
| | Power (re)distribution | Both designers and participants are asked to identify strategies for balancing power | The power of the designer and the participant is graphically represented, where the power of each stakeholder can vary from very low, 1, to very high, 5 |
| **Reflecting** | Collaborative decision-making | — | Documents each decision-making moment, identifying the participant's contribution to the project and the designer's knowledge, illustrating the combination of both for the final decision |
| | Diversity and inclusion | — | Lists the methods used to promote an inclusive project, which can be related to both the participants and the product/service being developed |
| | Moments of tension | — | Tensions are identified and designers are invited to reflect on the reasons why tensions occur and identify opportunities for improvement to prevent recurring tensions in the future |

Note: The sections marked with an * are optional

impact, reflected in Figure 6-*Power (re)distribution* in changes in power, especially with the participants.

By identifying "hidden" power dynamics between stakeholders and the moments they may emerge, there is greater awareness of possible imbalances. This can support the team to proactively address power dynamics.

*3) **Contextual factors**:* The 'Contextual factors' swimlane makes it easier to identify factors that can affect the power dynamics between designer and participant, highlighting which factors have the greatest impact on each phase; these factors can be cultural norms, social structures or even the location of activities with participants.

Figure 4 shows demographic and social factors identified as influencing power dynamics. However, these factors only had an impact on the 'define' and 'deliver' phases.

Examining the contextual factors that can affect power dynamics helps both designers and participants navigate the cultural, social and situational nuances that can affect collaboration. In our example, this mapping refers to contextual factors where the project is situated, this is not a required field.

*4) **External influences**:* This part of the map identifies external influences, such as economic constraints or regulations, and how they interact with power dynamics, which can potentially shape the design process.

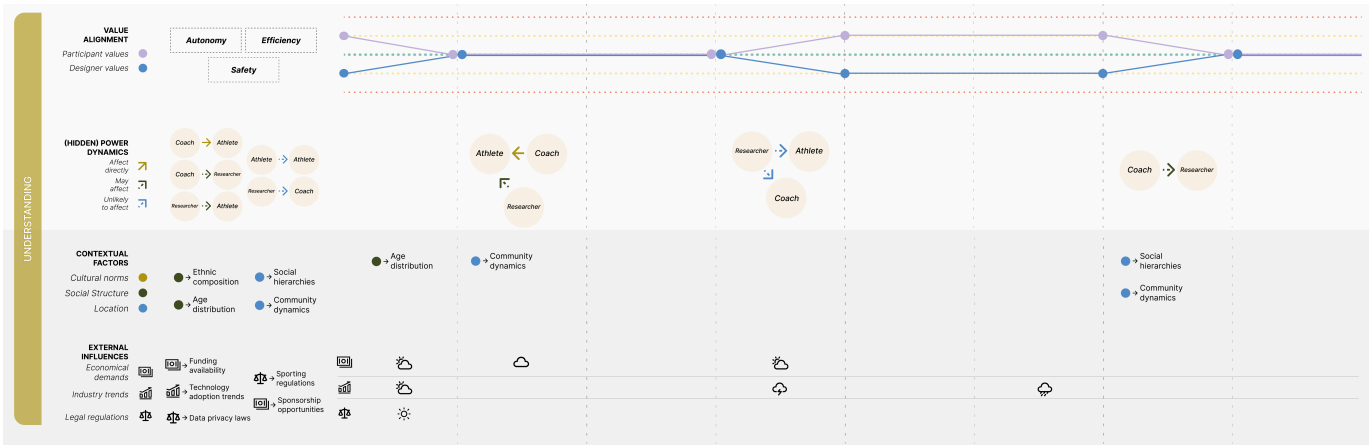

Fig. 4. Swimlanes of the *Understanding* section.

In this swimlane (Figure 4), economic and legal influences and industry trends stand out. Initially, the impact of these influences is assessed, revealing that technology adoption trends are not entirely favorable. Figure 4 shows that this influence intensifies over the course of the project, affecting the power dynamics considerably. As indicated in the 'Moments of tension' swimlane (Figure 6-*Moments of tension*), this influence undermines the power of the participants, causing their needs to be dismissed in favor of the use of technology.

By identifying external influences and how they can influence power dynamics, this field can help prepare the team to adapt to changes driven by external factors, improving their preparation in response to different situations. This field is optional, since the influences outside the project are also related to the context of the project.

### B. Mapping

This section of the journey map includes fields that contribute to a broader understanding of the design process, not only considering the direct interaction between designer and participant, but also the power of each involved party.

*1)* **Direct interaction moments**: This swimlane defines the design approach for the project, clarifying the expectations of the designers and participants in relation to the project. The process, the interactions, and moments of direct contact between the designer and the participant are identified for each phase of the project. Instances in which the designer and participant exchange knowledge and moments that constitute opportunities for empowerment are highlighted.

Considering that in our example, a co-design approach was adopted, the designer aims to create technological support to assist in the learning and performance improvement of pirouettes, while the participants seek to improve performance and accuracy (Figure 5). Moments of empowerment and knowledge exchange stand out, for example the 'co-creation session', where the designers can feature the contribution of the participants until that point. This mapping provides a comprehensive overview of the dynamics of collaboration between designer and participant.

*2)* **Power (re)distribution**: For this swimlane, at the moment of alignment, both designers and participants are asked to identify strategies for balancing power. Then, for each phase of the process, the power of the designer and the participant are graphically represented.

Figure 5 shows two methods for balancing power: the generation of collaborative ideas and the respect for the 'expertise' of the participants. As already mentioned, various factors can influence power, either positively or negatively. In this context, it is worth noting that the designer's power increases when external economic influence is favored (Figure 4-*External influences*).

The graphical representation of power helps to visualize the (un)balance of power between designers and participants, highlighting moments when power is or could be intentionally redistributed for a more equitable collaboration.

### C. Reflecting

The swimlanes in the 'Reflecting' section encourage pondering on power dynamics and the outcomes of the design process, after the activities took place. These swimlanes can also be used to further analyze the journey, namely the 'Moments of tension' swimlane, and to document situations that can be subject of reflection, specifying how these moments were dealt with. It should be noted that none of these fields include a moment of alignment, since this section is designed to reflect on the results of the design process.

*1)* **Collaborative decision-making**: This swimlane documents each decision-making moment, identifying the participant's contribution to the project as well as the designer's knowledge, illustrating the combination of both for the final decision. In this context, Figure 6-*Collaborative decision-making* shows that three decisions were made in collaboration with the participants, showing that even during the presentation and evaluation of the project's final product, there were opportunities for shared decisions.

By identifying the contribution of both parties, the roles of the participant and the designer in decision-making are emphasized and recognized, promoting true collaboration with the project's various parties.

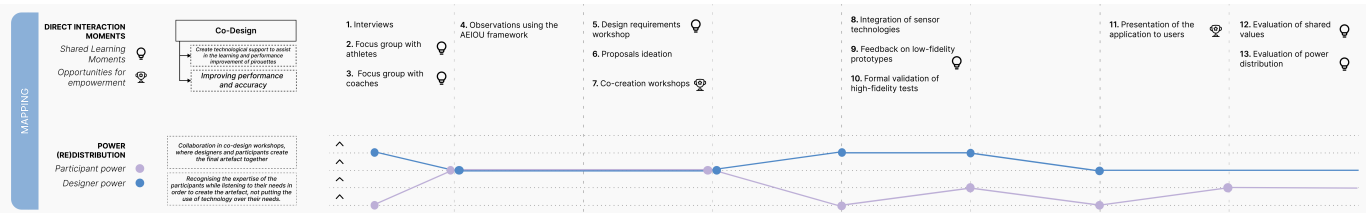

Fig. 5. Swimlanes of the *Mapping* section.

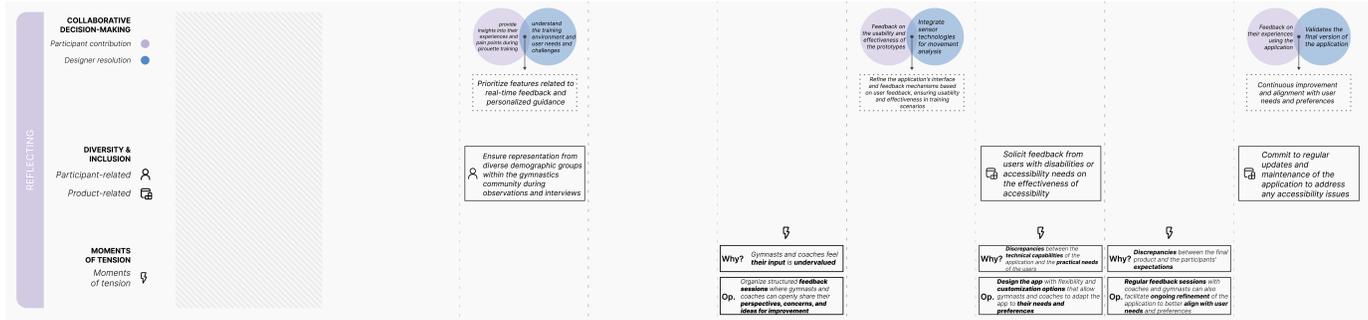

Fig. 6. Swimlanes of the *Reflecting* section.

*2) Diversity and inclusion*: This swimlane lists the methods used to promote an inclusive project, which can be related to both the participants and the product/service.

As illustrated in Figure 6-*Diversity and inclusion*, inclusion methods have been identified covering various phases of the design process. Some are directed at the participants, while others are focused on the end users and applied to the final product/service. In the 'discover' phase, the method applied focuses on ensuring the representation from diverse demographic groups within the gymnastics community during observations and interviews. By identifying inclusion methods, the integration of everyone is emphasized, and it is ensured that the design process is inclusive.

*3) Moments of tension*: This swimlane identifies tensions and designers are invited to reflect on the reasons why they occur and opportunities for improvement and for preventing them from reoccurring in the future are identified. Considering our example, three moments of tension were identified (Figure 6-*Moments of tension*): the first moment arises from a discrepancy between the technological capabilities of the application and the practical needs of gymnastics athletes. This moment of tension arises when the external factor relating to market trends is identified as detrimental to the project's power dynamics. The design team sees this as an opportunity to create customization options within the application to meet the athletes' needs. Identifying potential situations of tension and reflecting on their causes promotes a proactive approach to addressing challenges and minimizing tensions.

## V. CONCLUSIONS AND FUTURE WORK

The overarching goal of this work is to promote ethical and power-balanced design processes. This is an increasingly relevant subject in a world more and more characterized by the presence of technology, innovation and digital transformation. In developing these solutions, it is crucial that design and development teams are able to identify and navigate potential ethical issues, such as those arising from the management of power relations. This paper explored ethical dilemmas in Human-Computer Interaction (HCI) design and the strategies employed to address them. This knowledge informed the creation of the Participant-Designer Journey Map (PDJM), a practical tool to support teams in conducting an ethical and responsible design process.

Not having yet applied this tool in a real scenario is a limitation of this study, as is the fact that we relied on the feedback of only three experts for tool selection. Still, the selection of the proposal by experts provides a good indication of the potential interest of the PDJM. In addition, the PDJM results from a thorough methodological process of systematic exploration, anchored in the scientific literature reviewed and the analysis of existing tools. These guided each stage of the design and development process of the tool and provided a solid basis for the creation of the PDJM.

The example we used to elucidate the application of the tool highlights its practicality and effectiveness in pinpointing factors that could potentially induce tensions and foster a balanced distribution of power. This said, validation with different teams and projects is necessary and planned for future work. This validation will enable us to fully understand how designers, participants, and stakeholders experience the tool and determine how easy and effective the PDJM is in managing power imbalances. In the future, it would also be interesting to explore ways to manage potential instances where middle and/or top management may interfere with and/or overrule decisions made between those directly interacting with end-user participants and other relevant stakeholders. We should further seek to investigate the involvement of stakeholders who are not direct participants, but whose goals, needs, and preferences need to be taken into account, ranging from regulators to potential end users or customers/users of the participants' products or services.

The PDJM supports in the identification of power dynamics, creating the opportunity to anticipate and proactively act on potential emerging conflicts and power imbalances. In this way, the PDJM contributes to managing power dynamics between participants and designers across the design process. The PDJM offers a holistic approach to dealing with power dynamics, promoting an inclusive design process where the voices and needs of participants are valued and reinforced. This tool contributes to promoting ethical processes in HCI and the design and development of ethical and responsible solutions in this period of digital transformation.

ACKNOWLEDGMENT

We acknowledge the experts' contribution to validating the PDJM and the support of the Foundation for Science and Technology, project CISUC — UID/CEC/00326/2020.

REFERENCES

[1] C. Frauenberger, M. Rauhala, and G. Fitzpatrick, "In-Action Ethics," *Interacting with Computers*, vol. 29, no. 2, pp. 220–236, Jun 2016, doi: 10.1093/iwc/iww024.
[2] S. Harrison, D. Tatar, and P. Sengers, "The three paradigms of HCI," in *CHI '07: Proceedings of the SIGCHI Conference on Human Factors in Computing Systems*. New York, NY, USA: Association for Computing Machinery, Jan 2007.
[3] C. Frauenberger, "Entanglement HCI The Next Wave?" *ACM Transactions on Computer-Human Interaction*, vol. 27, no. 1, pp. 2:1–2:27, Nov 2019, doi: 10.1145/3364998.
[4] M. Malik and M. M. Malik, "Critical Technical Awakenings," *Journal of Social Computing*, vol. 2, no. 4, pp. 365–384, Dec 2021, doi: 10.23919/JSC.2021.0035.
[5] L. Tejo, "Ethics in the design process: from theory to practice," Master's thesis, Dept. Inf. Eng., Univ. of Coimbra, Coimbra, PT, September 2023, available at https://estudogeral.uc.pt/handle/10316/110511.
[6] C. Dindler, P. G. Krogh, K. Tikaer, and P. Norregard, "Engagements and Articulations of Ethics in Design Practice," *International Journal of Design*, vol. 16, no. 2, pp. 47–56, 2022, doi: 10.57698/v16i2.04.
[7] S. S. Chivukula, I. Obi, T. V. Carlock, and C. M. Gray, "Wrangling Ethical Design Complexity: Dilemmas, Tensions, and Situations," in *Companion Publication of the 2023 ACM Designing Interactive Systems Conference*, ser. DIS '23 Companion. New York, NY, USA: Association for Computing Machinery, Jul 2023, pp. 179–183, doi: 10.1145/3563703.3596632.
[8] R. Y. Wong, "Tactics of Soft Resistance in User Experience Professionals' Values Work," *Proceedings of the ACM on Human-Computer Interaction*, vol. 5, no. CSCW2, pp. 355:1–355:28, 2021, doi: 10.1145/3479499.
[9] S. Lindberg, P. Karlström, and S. Männikkö Barbutiu, "Design Ethics in Practice - Points of Departure," *Proceedings of the ACM on Human-Computer Interaction*, vol. 5, no. CSCW1, pp. 130:1–130:19, 2021, doi: 10.1145/3449204.
[10] S. S. Chivukula, C. R. Watkins, R. Manocha, J. Chen, and C. M. Gray, "Dimensions of UX Practice that Shape Ethical Awareness," in *Proceedings of the 2020 CHI Conference on Human Factors in Computing Systems*, ser. CHI '20. New York, NY, USA: Association for Computing Machinery, 2020, pp. 1–13, doi: 10.1145/3313831.3376459.
[11] C. M. Gray and S. S. Chivukula, "Ethical Mediation in UX Practice," in *Proceedings of the 2019 CHI Conference on Human Factors in Computing Systems*, ser. CHI '19. New York, NY, USA: Association for Computing Machinery, 2019, pp. 1–11, doi: 10.1145/3290605.3300408.
[12] C. Munteanu, H. Molyneaux, W. Moncur, M. Romero, S. O'Donnell, and J. Vines, "Situational Ethics: Re-thinking Approaches to Formal Ethics Requirements for Human-Computer Interaction," in *Proceedings of the 33rd Annual ACM Conference on Human Factors in Computing Systems*, ser. CHI '15. New York, NY, USA: Association for Computing Machinery, 2015, pp. 105–114, doi: 10.1145/2702123.2702481.
[13] B. Friedman, P. H. Kahn, A. Borning, and A. Huldtgren, "Value Sensitive Design and Information Systems," in *Early engagement and new technologies: Opening up the laboratory*, N. Doorn, D. Schuurbiers, I. van de Poel, and M. E. Gorman, Eds. Dordrecht: Springer Netherlands, 2013, vol. 16, pp. 55–95, doi: 10.1007/978-94-007-7844-3_4.
[14] O. T. D. Tarrant, J. Maddisom, "The data ethics canvas, the odi," Available at https://theodi.org/insights/tools/the-data-ethics-canvas-2021/, Jun 2021.
[15] A. C. W. Reijers, K. Levacher, "The ethics canvas," Available at https://www.ethicscanvas.org/canvas/index.php, Jun 2021.
[16] S. S. Chivukula, Z. Li, A. C. Pivonka, J. Chen, and C. M. Gray, "Surveying the Landscape of Ethics-Focused Design Methods," Aug 2022, doi: 10.48550/arXiv.2102.08909.
[17] T. Bratteteig and I. Wagner, "Disentangling power and decision-making in participatory design," in *Proceedings of the 12th Participatory Design Conference: Research Papers - Volume 1*, ser. PDC '12. New York, NY, USA: Association for Computing Machinery, 2012, pp. 41–50, doi: 10.1145/2347635.2347642.
[18] Y. Guo and D. G. Hoe-Lian, ""We Want to Hear Your Voice": Power Relations in Participatory Design," in *2014 11th International Conference on Information Technology: New Generations*. Las Vegas, NV, USA: IEEE, Apr 2014, pp. 561–566, doi: 10.1109/ITNG.2014.9.
[19] F. Tomasini Giannini and I. Mulder, "Towards a Power-Balanced Participatory Design Process," in *Proceedings of the Participatory Design Conference 2022 - Volume 2*, ser. PDC '22, vol. 2. New York, NY, USA: Association for Computing Machinery, 2022, pp. 111–117, doi: 10.1145/3537797.3537819.
[20] S. Erete, Y. Rankin, and J. Thomas, "A Method to the Madness: Applying an Intersectional Analysis of Structural Oppression and Power in HCI and Design," *ACM Transactions on Computer-Human Interaction*, vol. 30, no. 2, pp. 24:1–24:45, 2023, doi: 10.1145/3507695.
[21] C. Harrington, S. Erete, and A. M. Piper, "Deconstructing Community-Based Collaborative Design: Towards More Equitable Participatory Design Engagements," *Proceedings of the ACM on Human-Computer Interaction*, vol. 3, no. CSCW, pp. 216:1–216:25, Nov 2019, doi: 10.1145/3359318.
[22] A. Gautam and D. Tatar, "Empowering Participation Within Structures of Dependency," in *Proceedings of the Participatory Design Conference 2022 - Volume 1*, ser. PDC '22. New York, NY, USA: Association for Computing Machinery, 2022, pp. 75–86, doi: 10.1145/3536169.3537781.
[23] B. Brown, A. Weilenmann, D. McMillan, and A. Lampinen, "Five Provocations for Ethical HCI Research," in *Proceedings of the 2016 CHI Conference on Human Factors in Computing Systems*, ser. CHI '16. New York, NY, USA: Association for Computing Machinery, 2016, pp. 852–863, doi: 10.1145/2858036.2858313.
[24] M. Thinyane, K. Bhat, L. Goldkind, and V. K. Cannanure, "Critical participatory design: reflections on engagement and empowerment in a case of a community based organization," in *Proceedings of the 15th Participatory Design Conference: Full Papers - Volume 1*, ser. PDC '18. New York, NY, USA: Association for Computing Machinery, 2018, pp. 1–10, doi: 10.1145/3210586.3210601.
[25] S. Ballard, K. M. Chappell, and K. Kennedy, "Judgment Call the Game: Using Value Sensitive Design and Design Fiction to Surface Ethical Concerns Related to Technology," in *Proceedings of the 2019 on Designing Interactive Systems Conference*, ser. DIS '19. New York, NY, USA: Association for Computing Machinery, Jun 2019, pp. 421–433, doi: 10.1145/3322276.3323697.
[26] D. Oitzinger, "My favorite ideation exercise: 6–8–5 sketching, medium," Available at https://medium.com/@doitzi/my-favorite-ideation-exercise-6-8-5-sketching-76272205cd6f, Mar 2020.